\documentclass[aps,prl,twocolumn,showpacs,superscriptaddress,groupedaddress]{revtex4}  
\usepackage{graphicx}  
\usepackage{dcolumn}   
\usepackage{bm}        
\usepackage{amssymb}   
\usepackage{cases}
\usepackage{subfigure}
\begin{document}

\widetext

  \centerline{\em D\O\ INTERNAL DOCUMENT -- NOT FOR PUBLIC
DISTRIBUTION}


\title{Physical Origin of Nonlinear transport in organic
semiconductor
 at high carrier densities}
\author{Ling Li, Nianduan Lu, and Ming Liu }
\email{lingli@ime.ac.cn, liuming@ime.ac.cn}
\altaffiliation{Institute of Microelectronics, Chinese Academy of
Sciences, Beijing, 100029, China}

\date{\today}

\begin{abstract}
The charge transport in some organic semiconductors demonstrates
nonlinear properties and further universal power-law scaling with
both bias and temperature. The physical origin of this behavior is
investigated here using variable range hopping theory. The results
shows, this universal power-law scaling can be well explained by
variable range hopping theory . Relation to the recent experimental
data is also discussed.
\end{abstract}

\maketitle


Understanding of the charge transport mechanism in disordered
organic semiconductors such as conjugated and molecularly doped
polymers, is of crucial importance to the design and synthesis of
better materials. In organic semiconductors, due to the presence of
disorder, charge carriers are usually localized over spatially and
generically distributed transport sites. It is widely accepted that,
in such system, electrical conduction occurs by hopping, i.e.,
thermal assisted tunneling of charge carriers between localized
states. The hopping conductivity $\sigma$, is therefore decried as

\begin{equation}
\sigma\propto\exp\left(-\frac{E_A}{k_BT}\right).
\end{equation}
where the barrier $E_A$ (activation energy) arises due to disorder
between sites and due to nuclear polarization, and $k_B$ is the
Boltzmann constant. According to (1), the conductivity is expected
to vanish when the temperature approaches absolute zero. Recent
experiment on conductivity in the conjugated polymer
poly(2,5-bis-(3-tetradecylthiophen-2-yl)thieno[3,2-b] thiophene)
(PBTTT) in high carrier density field-effect transistors, however,
have demonstrated that the conductivity at low temperature is finite
\cite{heeger}. This unusual behavior has also been observed in other
organic semiconductors such as poly-3,4-ethylenedioxythiophene
stabilized with poly- 4-styrenesulphonic acid (PEDOT:PSS), rr-P3HT
\cite{leeuw1,leeuw2,heeger2}, and poly 3-hexylthiophene and 6,13-bis
triisopropyl-silylethynyl TIPS pentacene \cite{worne}. Moreover,the
charge transport shows universal power-law scaling with both bias
and temperature, more exactly, current $J\propto T^{\alpha}$ at low
voltages ($k_B T>eV$) and $J\propto V^{\beta}$ at low temperatures
($eV>k_BT$). Furthermore, when the scaled current density
$J/T^{1+\alpha}$ is plotted as a function of $eV/k_BT$, a universal
curve is obtained described by \cite{luttinger1,luttinger2}
\begin{equation}
J=J_0T^{1+\alpha}\sinh\left(\gamma'\frac{eV}{k_BT}\right)\mid\Gamma\left(\frac{1+\beta}{2}+i\gamma\frac{eV}{k_BT}\right)\mid^2.
\end{equation}
where the parameter $\alpha$ is derived from the measurements, $J_0$
and $\gamma$ are two fit parameters, $e$ is the elementary charge,
and $\Gamma$ is the Gamma function. The fit parameter $\gamma^{-1}$
has been related to the number of tunnel barriers between the
contacts and determines a crossover from Ohmic behavior to a
power-law dependence. This universal scaling law, is furtherly
ascribed to the nonlinear transport phenomena, i.e., $\sigma=I(V,
T)/V$ have a (stretched) exponential behavior, have been reported in
these materials. It was argued that, these observations are against
the classic hopping theory, and has been interpreted as a
fingerprint of Luttinger liquid behavior originating from one
dimensional transport in conjugated polymers. However, it has been
pointed out there is a problem with this interpretation. The actual
calculations \cite{heeger1} within the Luttinger model give
$\alpha=\beta$ and $\gamma=\gamma' $, which is not always consistent
with the parameters of the empirical fits \cite{fogler}. It has been
speculated that \cite{fogler}, in quasi-one-dimensional (1D) systems
the conventional VRH theory, at a low temperature regime, will
leadto this nonlinear behavior, which is because at low enough $T$,
the hopping length is not much smaller than the length $L$ of the 1D
wires. In this case, the hopping conductance deviates from the usual
formula (1). However, ont only in 1d system, but in 2D organic
semiconductors, this nonlinear behavior is also addressed
\cite{leeuw1,leeuw2}. Therefore, the model presented in  is possible
not the real physical origin of the nonlinear behavior. More
important, the material disorder, the main feature of disordered
organic semiconductor, has never been included in any
physical explanation.  \\
In this Letter, we show that the observed temperature dependent
conductivity or nonlinear behavior in organic semiconductors can be
consistently modeled, without invoking additional conduction
mechanisms, by considering that electrons may use variable range
hopping for conduction, the contribution of downwards hopping and
electric field compensated thermal activation will Simultaneously
lead to this behavior. More remarkable, We calculated the
macroscopic current based on the variable range hopping theory. The
current shows a power-law dependence on both temperature and
voltage. The renormalized current-voltage characteristics of various
polymers and devices at all temperatures collapse on a single
universal curve.\\
 \textit{Model}.---In general, the basis for models describing the charge
 transport in disordered semiconductors is Miller-Abrahams expressions
\cite{miller}, where the hopping transport takes place via tunneling
between an initial state $i$ and a target state $j$.The tunneling
process is described by

\begin{equation}
  \nu=\nu_0\exp\left(-u\right)=\nu_0\left\{
   \begin{array}{c}
   \exp\left(-2\alpha
R_{ij}-\frac{E_j-E_i}{k_BT}\right),  E_i>E_j\\
  \left(-2\alpha R_{ij}\right).  \qquad\qquad\qquad E_i<E_j\\
   \end{array}
  \right.
  \end{equation}
Here, $\nu_0$ is the attempt-to-jump frequency, $R_{ij}$ is the
hopping distance, $u$ is the hopping range \cite{arkhipov2},$E_i$
and $E_j$ are the energies at sites $i$ and $j$, respectively, and
$\alpha$ is the inverse localized length. However, in real organic
semiconductor systems, when an electric field $F$ exists, this
electric field will lower the Coulomb barrier, which leads to a
reduction in the thermal activation energies, and the hopping range
with normalized energy ($\epsilon=E/k_BT$) can therefore be
rewritten as \cite{apsley,li3}
\begin{equation}
 u=\left\{
   \begin{array}{c}
 2\alpha\left(1+\beta\cos\theta\right)
R_{ij}+\epsilon_j-\epsilon_i,  \epsilon_j>\epsilon_i-\beta\cos\theta\\
  2\alpha R_{ij}. \qquad\quad \qquad\qquad\qquad \epsilon_j<\epsilon_i-\beta\cos\theta\\
   \end{array}
  \right.
\end{equation}
where $\beta=Fe/{2\alpha k_BT}$ and $\theta$ is the angle between
$R_{ij}$ and the electric field ranging from $0$ to $\pi$. For a
site with energy $\epsilon_i$ in the hopping space, the most
probable hop for a carrier on this site is to an empty site at a
range $u$, for which it needs the minimum energy. The conduction is
a result of a long sequence of hops through this hopping space. The
average hopping range $R_{nn}$ can be obtained following the
approach used our previous work \cite{li3}, the mobility at energy
$\epsilon_i$ is
\begin{equation}
\mu\left(\epsilon_i,T,\beta\right)=\frac{\nu_0}{F}\bar{x_f}\exp\left(-R_{nn}\right)
\end{equation}
where $\bar{x}_f$ is the average hopping distance along the electric
field \cite{apsley}. The total conductivity of the hopping system is
\begin{equation}
\sigma\left(T,\beta\right)=\int_{-\infty}^{\infty}eg\left(\epsilon_i\right)f\left(\epsilon_i,\epsilon_F\right)\mu\left(\epsilon_i,T,\beta\right)k_BTd\epsilon_i.
\end{equation}
Where $g\left(\epsilon\right)$ is the density of states,
$f\left(\epsilon_i,\epsilon_F\right)=1/\left(1+\exp\left(\epsilon_i-\epsilon_F\right)\right)$
is the Fermi-Dirac distribution with $\epsilon_F$ denoting the Fermi
level. We take the Gaussian form of
 $g\left(\epsilon\right)=\frac{N_t}{\sqrt{2\pi}\sigma_0}\exp\left(-\frac{\epsilon^2}{2\sigma_0^2}\right)$
as the DOS in the full manuscript \cite{bassler5}, where $N_t$ is
the number of states per unit volume and $\sigma_0=\sigma'/kT$
indicates the width of the DOS. $N_t=1\times 10^{28}m^{-3}$ is used
in the full manuscript as a typical value for the relevant organic
semiconductor. Please note the experimental data in \cite{heeger} is
preformed in organic thin film transistor, in this situation, the
Fermi level is controlled by the gate voltage, and can be calculated
as the work in \cite{li2}. Based on equation(6), the temperature
dependence of the calculated conductivity is calculated, as shown in
the blue line of Fig.
 1. The input parameters are $F=9\times 10^6 V/m$, $\alpha^{-1}=4.1\AA$,
 $\sigma_0=0.09$eV, $\nu_0=1.7\times 10^{12} sec^{-1}$, and the gate voltage $V_g$ of organic thin film
 transistor is $150V$. It is shown that our model gives a crossover of
the conductivity with decreasing temperature. Above the crossover
temperature, the conductivity has an activated behavior, whereas
below this temperature, the conductivity depends very weakly on
temperature. Good agreement between calculation and experimental
data is obtained.\\

\begin{figure}[h]
             \centering \scalebox{0.35}{\includegraphics{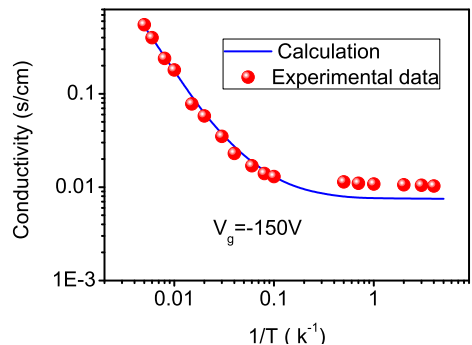}}
             \vspace*{-7pt}\caption{Computed conductivity as a function of the
             temperature. The dots presents the experimental data from \cite{heeger}.}
             \end{figure}
To calculate the macroscopic current using the equation (6),the
applied potential $V$ enters via charge carriers to traverse under
the field $F=V/L$, where $L$ and $W$ are the distance between the
electrodes and the width of device, respectively. The current $I$ is
given by
\begin{equation}
J=\frac{W\int_0^V\sigma(V')dV'}{L}.
\end{equation}
The calculated current curves at temperatures between 10 and 50 K
are presented on a double logarithmic scale in Fig. 2. The
parameters chosen here are the same as in Fig. 1. It is found here,
the transport at high temperature is Ohmic, and linear in bias, at
all voltages. The current decreases with decreasing temperature, and
at low temperature, the output curves become non-linear. The
transition voltage from linear to superlinear behavior decreases
with decreasing temperature. This conclusion is consistent with
experimental observation.
\begin{figure}[h]
             \centering \scalebox{0.35}{\includegraphics{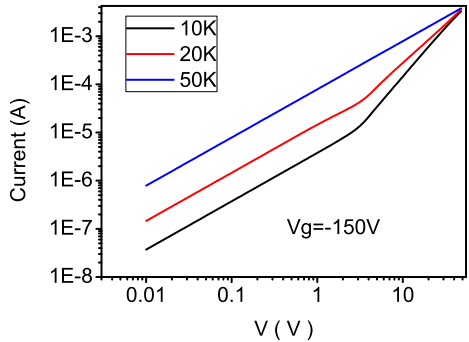}}
             \vspace*{-7pt}\caption{Computed current as a function of the
             temperature and voltage.}
             \end{figure}

 In fact, it has been pointed out,  any weakly temperature dependent data set that resembles a power-law can
be made to fit onto a single line if plotted in this way with an
appropriate choice of parameters. Therefore, the nonlinear behavior
of conductivity, should be response for the universal power law
scaling.  Therefore,  let us interpret the nonlinear charge
transport in organic semiconductors, i.e., temperature independent
conductivity at low temperature regime in the next step.\\
Actually, temperature independent conductivity has been discussed in
different disordered materials
\cite{vrh1,vrh2,dna,vrh3,vrh4,vrh5,vrh6}. The physical reason has
been attributed to the phonon emission at low temperature
\cite{vrh1}, thermal structural fluctuations in disordered materials
localize electronic wave functions, giving rise to a
temperature-dependent localization length \cite{dna}, or weak
Coulomb blockade \cite{vrh3,vrh4}. In what follows, we thoroughly analyze the above formulation.\\
According to equation (4), the hopping probability depends on both
the spatial and energetic separation of the hopping sites, it is
natural to describe the hopping processes in a four-dimensional
hopping space, with three spatial coordinates and one energy
coordinate. In this hopping space, a range $u$ given by the
magnitude of the exponent in equation 3, represents a distance in
four-dimensional hopping space, indicating the hopping probability.
\begin{figure}[h]
             \centering \scalebox{0.35}{\includegraphics{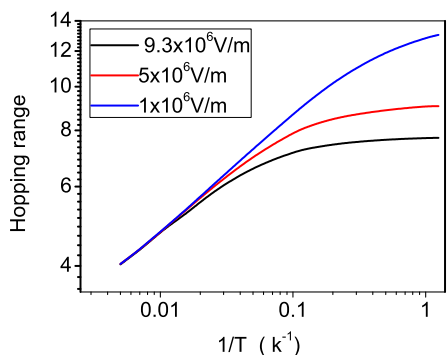}}
             \vspace*{-7pt}\caption{Computed hopping range from the Fermi level as a function of the
             temperature and electric field. }
             \end{figure}

Fig. 3 shows the temperature dependence of the hopping range under
different electric field. The parameters used here are the same as
in Fig. 1. We clearly identify a weakly temperature dependent
hopping range for low temperatures and  large electric field. To
investigate the physical origin of this behavior, let us obtain the
average hopping range for carrier at energy $\epsilon_i$  from
\cite{mott,arkhipov1,arkhipov2,baranovskii1,baranovskii2,li1}
\begin{eqnarray}
N\left(\epsilon_i,R',\beta\right)=1
\end{eqnarray}
where $N\left(\epsilon_i, R', \beta\right)$ is the finial states
enclosed by the contour $u$ as
\begin{eqnarray}
N\left(\epsilon_i,R',\beta\right)
=\nonumber\qquad\qquad\qquad\qquad\qquad\qquad\qquad\qquad\qquad\\
\frac{1}{8\alpha^3}\int_0^{\pi}d\theta\sin\theta\int_0^{R'} dr2\pi
R'^2\int_{-\infty}^{R'+\epsilon_i-r\left(1+\beta\cos\theta\right)}d\epsilon
\nonumber\\\times
g\left(\epsilon\right)\left[1-f\left(\epsilon,\epsilon_F\right)\right].
\end{eqnarray}
After changing variables, equation (9) is rewritten as rewritten as
equation (10) (see the top of next page), where
$\lambda\left(\epsilon\right)=g\left(\epsilon\right)\left[1-f\left(\epsilon,\epsilon_F\right)\right]$.
Based on this equation, the hopping event can be dived into three
parts: downwards hopping, thermal activated hopping and field
direction hopping. For some hopping range $u$ obtained by solving
equation (8), we can separate the contribution of these three parts,
as shown in Fig. 4. At high temperature, the thermal activated
hopping is dominant; For the low temperature, however, the hopping
event is totally determined by field direction hopping. Therefore,
for the low temperature, we approximate  the  equation (10) as

\begin{table*}[t]
\begin{eqnarray}
N\left(\epsilon_i,R',\beta\right)\propto\underbrace{\frac{2R'^3}{3}
\int_{-\infty}^{\epsilon_i+R'}\lambda\left(\epsilon\right)d\epsilon}_{downwards
hopping}+\underbrace{\frac{2}{3}\int_{\epsilon_i-\beta
R'}^{\epsilon_i+R'}\lambda\left(\epsilon\right)\frac{\left(R'+\epsilon_i-\epsilon\right)^3}{\left(1+\beta\right)^3}d\epsilon}_{thermal
activated
hopping}\nonumber\qquad\qquad\qquad\qquad\qquad\qquad\qquad\qquad\qquad\qquad\\+\underbrace{
\left\{ {\begin{array}{*{20}c}
 \frac{1}{3}\int_{\epsilon_i-\beta R'}^{\epsilon_i+R'}\lambda\left(\epsilon\right)\left(R'^3-\left(\frac{R'+\epsilon_i-\epsilon}{1+\beta}\right)^3\right)d\epsilon-
 \frac{1}{3}\int_{\epsilon_i+\beta R'}^{\epsilon_i+R'}\lambda\left(\epsilon\right)\left(R'^3-\left(\frac{R'+\epsilon_i-\epsilon}{1-\beta}\right)^3\right)d\epsilon,\beta<1 \\
  \frac{1}{3}\int_{\epsilon_i+
  R'}^{\epsilon_i+\beta R'}\lambda\left(\epsilon\right)\left(R'^3-\left(\frac{R'+\epsilon_i-\epsilon}{1-\beta}\right)^3\right)d\epsilon+
 \frac{1}{3}\int_{\epsilon_i-\beta R'}^{\epsilon_i+R'}\lambda\left(\epsilon\right)\left(R'^3-\left(\frac{R'+\epsilon_i-\epsilon}{1+\beta}\right)^3\right)d\epsilon,\beta>1 \\
\end{array}} \right.}_{field direction hopping}
\end{eqnarray}
\hrule
\end{table*}

\begin{equation}
N\left(\epsilon_i,R',\beta\right)\propto\int_{\epsilon_i}^{\epsilon_i+\left(1+\beta\right)R'}\left(R'^3-\left(\frac{\left(1+\beta\right)R'+\epsilon_i-\epsilon}{1+\beta}\right)^3\right)d\epsilon
\end{equation}
If the constant DOS $g$ is assumed, , equation (8) reduces to
\begin{equation}
N\left(\epsilon_i,R',\beta\right)=\frac{3g\pi
k_BT}{16\alpha^3}R'^4(1+\beta)\approx\frac{3g\pi
k_BT}{16\alpha^3}R'^4\beta=1
\end{equation}

In this situation, the hopping range
$R_{nn}=(\frac{8\alpha^2}{3\beta g\pi })^{1/4}$ is temperature
independent. Physically this means that the initial energy
difference ($(1+\beta)R'$ here) between the hopping states is
completely compensated by the energy gain of the charge carrier
hopping in the direction of the electric force. Since the energy
needed in the hopping process is fully provided by the electric
field, no thermal activation is required anymore and a field induced
tunneling current is dominant.

\begin{figure}[h]
             \centering \scalebox{0.35}{\includegraphics{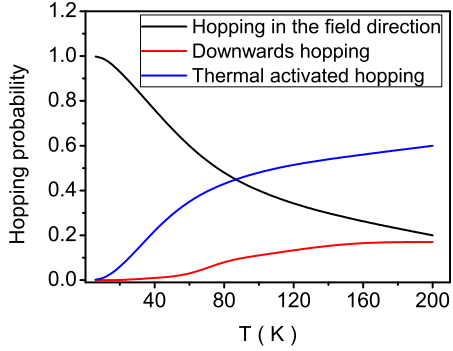}}
             \vspace*{-7pt}\caption{Computed hopping range from the Fermi level as a function of the
             temperature and electric field. }
             \end{figure}
Next, we want to analyze the universal power-law scaling of charge
transport with both bias and temperature, which has been repotted in
different organic semiconductors. According to equation (7) , we
present the scaled current $I/T^{1+\alpha}$, as a function of
relative energy $eV/k_BT$, in Fig.5. The parameters used for this
calculation are: $L=1\mu m$ and $W=100\mu m$, the other parameters
are the same as in Fig. 1. For different voltages and temperatures,
the scaled current collapse onto a single curve. At low values of
$eV/k_BT$, the scaled current increases linearly with relative
energy . At high values of $eV/k_BT$, the scaled current increases
superlinearly.Subsequently, the scaled current was fitted to
Luttinger liquid model of the one-dimensional metal (equation (2))
using the values of the parameters $\alpha$ of 1.1, $\gamma^{-1}$ of
300, and $\beta$ of 3. Figure 3 shows that an excellent fit is
obtained for the VRH model and Luttinger model.

\begin{figure}[h]
             \centering \scalebox{0.35}{\includegraphics{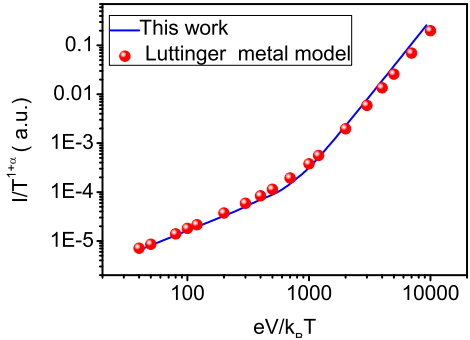}}
             \vspace*{-7pt}\caption{Comparison between the Scaled current calculated using VRH theory and Luttinger model.}
             \end{figure}
\label{compmetal} To address the reason for the power-law scaling
using VRH theory, the current density is approximated in Mott¡¯s
formalism by
 \begin{equation}
 J\approx
 2e{R_{nn}}k_BT\nu_0\exp\left(-R_{nn}\right)\sinh\left(\frac{eR_{nn}F}{k_BT}\right)
 \end{equation}
If the contribution of $\frac{eF}{k_BT}$ to the hopping range is
small, equation (13) is approximated as
\begin{eqnarray}
 \frac{J}{e{R_{nn}}k_BT\nu_0} \approx 2R_{nn}\frac{e F}{k_BT}
 \end{eqnarray}
The hopping range $R_{nn}\propto T^{-{1/4}}$ in the high temperature
\cite{temperature1,temperature2}, therefore, the scaling low
$\frac{J}{T^{1+1/2}}\propto\frac{eV}{k_BT}$ can be easily obtained
from equation (14).At even higher electric field, the carrier will
more possible transport along the field and the energy difference is
determined by $eFR_{ij}$, the hopping range is actually temperature
independent as $R_{nn}\propto F^{-1/2}$ \cite{field1,field2},
equation (14) can be approximated as
\begin{equation}
\frac{J}{T^{1+1/2}}\propto\left(\frac{eV}{k_BT}\right)^{-1/2}\exp\left(\frac{eV}{k_BT}\right)
\end{equation}
In this situation, $\frac{J}{T^{1+1/2}}$ will superlinear increases
with $eV/k_BT$. Please note, the real parameter $1/2$ is related to
energy disorder and carrier concentration, which has been observed
in \cite{heeger,heeger2}.

 In conclusion, we have shown that the classic VRH theory
leads to a unified description of the nonlinear transport
characteristics of semiconducting polymers at high carrier
densities. Furthermore, we showed that, in a single plot, the
calculated VRH current collapse and a universal curve is obtained,
which agreement with Luttinger model. It is demonstrated that the at
low temperature, the field tunneling is dominant and contributes to
the charge nonlinear transport. We further find that a scaling
approach is insufficient to test for charge transport mechanism in
organic semiconductors, as it shows apparent VRH will lead to this
scaling behavior.

 Financial support from NSFC (No. 60825403)
and National 973 Program 2011CB808404 is acknowledged.

\end{document}